\documentclass[12pt,a4paper]{revtex4}
\usepackage{amsfonts}
\usepackage{color}
\usepackage{eepic}
\usepackage{epic}
\usepackage{ulem}
\usepackage{graphicx}
\usepackage{texdraw}
\usepackage{epsfig}
\newcommand{\muMS}{\bar\mu_{{\rm MS}}}
\begin{document}
\title{Viscosity and thermodynamic properties of QGP in relativistic heavy ion collisions}
\author{ Vinod Chandra$^{a}$}
\email{vinodc@iitk.ac.in}
\author{ V. Ravishankar $^{a,b}$}
\email{vravi@iitk.ac.in, vravi@rri.res.in}
\affiliation{$^{a}$ Department of Physics, Indian Institute of Technology 
Kanpur, UP, India, 208 016}
\affiliation{$^{b}$ Raman Research Institute, C V Raman Avenue, Sadashivanagar, 
Bangalore, 560 080, India}
\date{\today}
\begin{abstract}
We study the viscosity and thermodynamic properties  of QGP  at RHIC by employing the recently
extracted equilibrium distribution functions from two hot QCD equations of state of $O(g^5)$ and $O(g^6\ln(1/g))$ respectively. 
After obtaining the temperature dependence of  energy density, and entropy density,
we focus our attention on the determination of shear viscosity for a rapidly expanding interacting plasma, as a function of temperature. We find that interactions significantly decrease the shear viscosity. They decrease the  viscosity to entropy density ratio, $\eta/{\mathcal S}$ as well.
\end{abstract}
\maketitle
\vspace{2mm}
{\bf PACS}: 25.75.-q; 24.85.+p; 05.20.Dd; 12.38.Mh

\vspace{3mm}
{\bf Keywords:} Quark Gluon Plasma, hot QCD, shear viscosity, entropy density, equation of state, viscosity to entropy density ratio

\section{Introduction}
Recent experimental results from RHIC\cite{expt} reveal that the QGP produced in heavy ion collisions 
behaves like an almost perfect fluid with very low viscosity \cite{dtson,bair,drescher}. In belying the earlier expectations that the deconfined phase would show nearly ideal behavior at temperatures close to $T_c$, the results from the flow measurements signal that the deconfined phase is  strongly interacting. 
  Lattice studies \cite{fkarsch} also predict  that
the equation of state for QGP is about 10 percent away from ideal EOS even at $T\sim 4T_c$. 
Therefore, studies based on  the ideal EOS in the computations are certainly 
inadequate  to address the physics in this domain of QCD. 

In an attempt to understand the flow measurement results, a variety of techniques have been employed,
including  the employment of AdS/CFT correspondence. A strong result from these studies is a lower
bound on the viscosity to entropy density ratio given by $\eta/\mathcal{S} < \frac{1}{4\pi} \approx 0.08$ \cite{dtson}. 
Yet another approach is to study a strongly coupled classical plasma \cite{shuryak} which again yields a
value close to the lower bound mentioned above. There are two lattice results, one by Meyer \cite{meyer}
who obtains a value 0.13 for $\eta/\mathcal{S}$ in  pure gauge theory at $T =1.56 T_c$. The other
result, also in pure aguge theory is due to Nakamura \cite{nakamura} who finds that  $\eta/\mathcal{S} < 1$.
 Recall that the analyses
\cite{adare,gavin}  based on $v_2$ measurements \cite{expt1} arrive at values that vary from $0.08 - 0.2$.
Results of other  studies \cite{drescher,lacey,bair,greiner} also yield numbers in the same range. Interestingly, Asakawa et al \cite{asprl,asakawa,abhijit} find that the ratio can take a value which is
smaller than the lower bound set by AdS/CFT studies, depending on the value assigned to the transport
parameter $\hat{q}_R$.

It is noteworthy that while the perfect liquid picture implies a strongly interacting QGP (sQGP),
most of the estimates made above employ a perturbation about the ideal distribution for quarks and gluons.
While this could simplify matters, it is  important to find out  what an EOS which includes
the interactions would predict for $\eta/\mathcal{S}$. In doing so, one could also gain an appreciation
of the nature of interactions in sQGP. In this paper, we study the predictions of EOS which are based
on improved perturbative QCD \cite{zhai,kajantie}. The implications of lattice EOS will be taken up in a subsequent paper. Our study is based on  \cite{chandra1,chandra2} where it has been shown how
 the EOS may be adapted to study the properties of QGP in heavy ion collisions. It is worth mentioning that the above mentioned works have demonstrated the viability of the EOS since they yield reasonable values for the dissociation temperatures for $J/\Psi$ and $\Upsilon$.

The paper is organized as follows. In Section (II), we introduce the EOS based on pQCD and
review the work contained in 
\cite{chandra1,chandra2}. We proceed to determine the   the temperature dependence of thermodynamic observables(energy density and entropy density) in Section(III).  In Section (IV), We obtain the  expressions for anomalous and collisional contributions to the parton shear viscosities, in the presence of interactions. We further study the behavior of viscosity to entropy density($\eta/{\mathcal S}$) as a function of temperature for pure gauge theory plasma, and compare the
results with the ones obtained from an ideal gas distribution. In Section(VI), we conclude this work.

\section{Hot QCD equations of state and their quasiparticle description}
Recently Chandra {\it et. al}\cite{chandra1,chandra2} have considered two EOS based on pQCD, and developed a self-consistent method to recast them
 in terms of non-interacting/weakly interacting quasi particles with effective fugacities.  Since the method is employed here, we briefly review the work.

The  EOS which we label EOS2 \cite{kajantie} is given by
\begin{eqnarray}
\label{eq2}
P_{g^6\ln(1/g)}&=&\frac{8\pi^2}{45\beta^4}\bigg \lbrace (1+\frac{21N_f}{32})-\frac{15}{4}(1+\frac{5N_f}{12})\frac{\alpha_s}{\pi}
\nonumber\\ &&+30(1+\frac{N_f}{6})(\frac{\alpha_s}{\pi})^{\frac{3}{2}} 
+\bigg[(237.2+15.97N_f\nonumber\\ 
&&-0.413 N_f^2 +\frac{135}{2}(1+\frac{N_f}{6})\ln(\frac{\alpha_s}{\pi}(1+\frac{N_f}{6}))\nonumber\\
&&-\frac{165}{8}(1+\frac{5N_f}{12})(1-\frac{2N_f}{33})\ln[\frac{\muMS\beta}{2\pi}]\bigg](\frac{\alpha_s}{\pi})^2\nonumber\\
&&+(1+\frac{N_f}{6})^{\frac{1}{2}}\bigg[-799.2-21.99N_f-1.926N_f^2\nonumber\\
&&+\frac{495}{2}(1+\frac{N_f}{6})(1+\frac{2N_f}{33})\ln[\frac{\muMS\beta}{2\pi}]\bigg](\frac{\alpha_s}{\pi})^{\frac{5}{2}} 
 \nonumber\\
&&+\frac{8\pi^2}{45}T^4 \biggl[1134.8+65.89 N_f+7.653 N_f^2\nonumber\\
 &&-\frac{1485}{2}\left(1+\frac{1}{6} N_f\right)\left(1-\frac{2}{33}N_f\right)
\ln(\frac{\muMS}{2\pi T})\biggr]\nonumber\\
&&\times\left(\frac{\alpha_s}{\pi}\right)^{3}
(\ln \frac{1}{\alpha_s}+\delta)\bigg \rbrace.\nonumber\\
\end{eqnarray}

The other EOS \cite{zhai}, which we call EOS1,  is of $O(g^5)$, and is obtained from this equation by dropping   the last term which has contributions of $O(g^6(\ln(1/g)+\delta))$. The phenomenological parameter $\delta$ is introduced in \cite{kajantie} to incorporate the undetermined contributions
of $O(g^6)$. 

The above equation of state has several ambiguities, associated with the renormalization scale
$(\mu_{\bar MS})$, the scale parameter $\Lambda_T/\Lambda_{\bar{MS}}$ which occurs in the expression for the running copulping constant $\alpha_s$, and the value of the phenomenological parameter $\delta$.
The ambiguity associated with $(\mu_{\bar{MS}})$ has been discussed well in literature and a popular way out is the BLM criterion due to Brodsky, Lepage and Mackenzie \cite{blm}. In this criterion, which is chosen to make the highest power of $N_f$ vanish in the highest perturbative order, the value of $(\mu_{\bar{MS}})$ is allowed to  vary between $\pi T$ and 4$\pi T$ \cite{neito}. In this paper we choose the renormalization scale $\mu_{\bar MS}=2.15\pi T\approx 6.752 T$\cite{avrn1} close to the central value $2\pi T$. One feature of this particular choice is that all the contributions due to the logarithms containing $\mu_{\bar MS} $ are very small. For the scale parameter  $\Lambda_T$, we follow Huang and Lissia \cite{shaung} and set $\Lambda_T/\Lambda_{\bar MS}=\exp(\gamma_E + 1/22)/4\pi \approx 0.148$,
since with this choice, they find among other things  that the coupling $g^2(T)$ is optimal for lattice perturbative calculations. The same value has also been em[ployed by others. see e.g. \cite{kajantie,avrn1}.
Finally, we set   $\Lambda_{\bar{MS}} = T_c$, which is close to the value  $0.87T_c$ found by Gupta \cite{sgupta}.

We turn our attention to the phenomenological parameter $\delta$.  The optimal value of $\delta$ depends on the
choice of the renormalization scale and the order in which the running coupling constant is determined. Blaizot, Iancu and Rebhan \cite{rebh_2}
find that the optimal value is given by $\delta=1/3$ if one employs the two loop running coupling constant while, the one loop running coupling constant yields $\delta$ in the range 0.7-0.9 \cite{kajantie,chandra2}. In this paper, we find that $\delta$ in the range 0.8 to 1.2 yields the best fit with the lattice results.
The important point here is that once the phenomenological parameter $\delta$ is fixed by comparing  EOS2 with lattice EOS, it can be employed to study the properties of QGP. As regards EOS1, we note that the matching with the lattice results has been found to be merely qualitative \cite{chandra2}.

Let us briefly review  the underlying idea and the findings of recent two papers\cite{chandra1,chandra2}.
In Ref.\cite{chandra1}, it has been shown that the interaction effects in EOS1 and EOS2 can be captured in terms 
of the effective fugacities  ($z_g$,$z_q$) for the quasi gluons and quarks.  
The effective fugacities are determined self-consistently order by order. 
The mapping has been found to be accurate up to about 5\% error. Therefore, we expect an error of the 
same order for all the quantities which can directly be derived from the pressure  for eg., the energy-density and the entropy density.

 Interestingly,
 the temperature dependence
of the screening length which is subsequently determined  is seen to  qualitatively agree with the lattice  results of Zantow\cite{zantow}.
In Ref.\cite{chandra2}, the quasi-particle description developed in  Ref.\cite{chandra1} has been combined with the 
formulation of the response function of QGP\cite{akr}, and the dissociation temperatures for  $J/\Psi$ and $\Upsilon$
have been estimated. These numbers are again reasonably close to the predictions of other theoretical works\cite{satz,prd75}.
This motivates us to further utilize EOS1 and EOS2 to study the behavior of 
thermodynamic quantities such as energy density, entropy density, and most importantly, the transport parameters, shear viscosity $\eta$ and viscosity to entropy density ratio $\eta/{\mathcal S}$, for the rapidly expanding plasma. In addressing this, we
 generalize the recent work of Asakawa, M\"uller and Bass\cite{asakawa} on the transport properties of interacting QGP. 

\section{ Thermodynamic observables}
Let us now turn our attention to study the behavior of thermodynamic observables. We consider energy density ($\epsilon$) first which brings out the physics of the quasi-particle description manifestly. We then 
determine the entropy density(${\mathcal S}$), for both
EOS1 and EOS2. We principally employ the method developed in \cite{chandra1}. 

As mentioned, EOS1 and EOS2 are mapped to the corresponding equilibrium functions with the quarks and
the gluons possessing effective fugacities:
\begin{equation}
\label{eq3a}
f^{g/q}_{eq}=\frac{1}{\bigg[z^{-1}_{g/q}\exp(\beta p)\mp 1\bigg]}
\end{equation} 
where all the interaction effects are captured in the fugacities 
 $z_{g/q}\equiv \exp{\beta\mu_{g/q}}$. The form of $z_{g/q}$ as a function of
temperature is given in \cite{chandra2}. With these distributions, it is straight forward
to determine the thermodynamic quantities.
\subsection{The energy-density}
Notwithstanding appearances, the energy of the 
quasi-gluons and quasi-quarks in not merely given by the relation $E_p=p$.  Rather, it should be determined from
the fundamental thermodynamic relation between the energy density and the partition function,
$\epsilon=-\partial_\beta \ln(Z)$. Substituting for the partion function in terms of quasi-gluons and 
quasi-quarks we obatin,
\begin{equation}
\label{eq3}
\epsilon_{q/g}=\frac{\nu_{g/q}}{8\pi^3}\int (p+T^2\partial_T \ln (z_{g/q}))f^{g/q}_{eq},
\end{equation}
where $\nu \equiv(\nu_g,\nu_q)= (2(N^2_c-1),4N_cN_f)$.
The modified dispersion relation reads,
\begin{equation}
\label{des}
 E_p=p+T^2\partial_T \ln (z_{g/q}).
\end{equation}

After performing the momentum integration in Eq.\ref{eq3}, we obtain the following
expression for the energy-density:
\begin{eqnarray}
\label{eq4}
\frac{\epsilon}{T^4}&=&\frac{\nu_g}{2\pi^2} 6{\mathcal PolyLog}[4,z_g)]
           - \frac{\nu_q}{2\pi^2} 6 {\mathcal PolyLog}[4,-z_q)]\nonumber\\
&&+\frac{(\Delta_g +\Delta_q)}{T^4},
\end{eqnarray}

where
\begin{eqnarray}
 \Delta_g&=&T^2\partial_T \ln(z_g){\mathcal N_g}\nonumber\\
 \Delta_q&=&T^2\partial_T\ln(z_q){\mathcal N_q},
\end{eqnarray} are the contributions
from the quasi-gluons and quasi-quarks to the trace anomaly. The second term in the dispersion relation Eq.(\ref{des}) may be thus looked upon as the anomalous component of the dispersion relation. 
The quantities ${\mathcal N_g}$ and ${\mathcal N_q}$ are the quasi-gluon and quasi-quark number densities and having the following 
form,
\begin{eqnarray}
{\mathcal N_g}=\frac{\nu_g T^3}{\pi^2} {\mathcal PolyLog}[3,z_g]\nonumber\\
{\mathcal N_q}=-\frac{\nu_q T^3}{\pi^2}{\mathcal PolyLog}[3,-z_q].
\end{eqnarray}

\begin{figure}[htb]
\label{fig1}
\vspace*{-70mm}
\hspace*{-40mm}
\psfig{figure=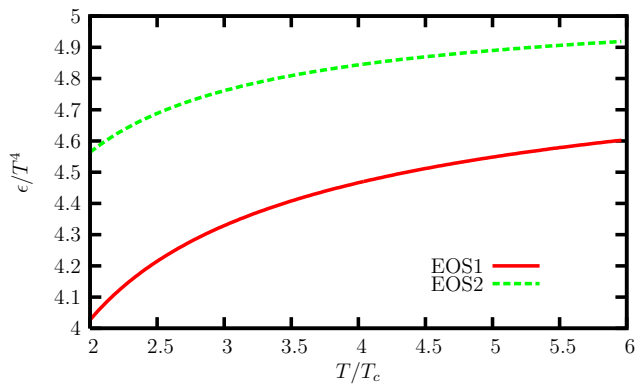,width=140mm}
\vspace*{-80mm}
\caption{Behavior of the Energy density for pure gauge theory as a function of temperature}
\end{figure}

\begin{figure}[htb]
\label{fig2}
\vspace*{-70mm}
\hspace*{-40mm}
\psfig{figure=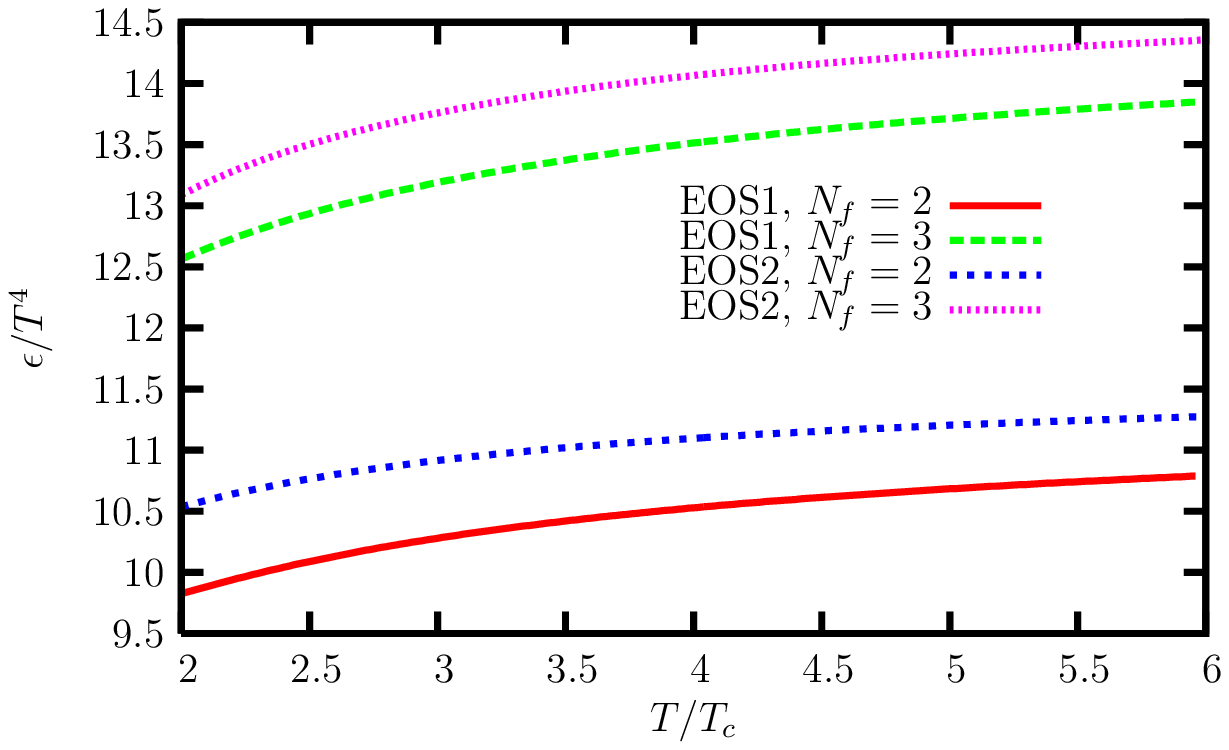,width=140mm}
\vspace*{-80mm}
\caption{Behavior of the Energy density in 2- and 3-flavor QCD as a function of temperature}
\end{figure}

\begin{figure}[htb]
\label{fig3}
\vspace*{-70mm}
\hspace*{-40mm}
\psfig{figure=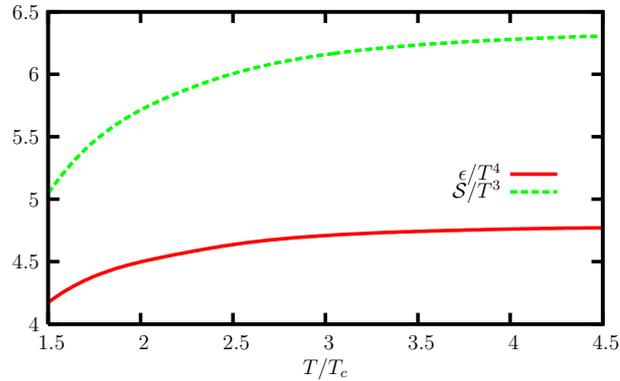,width=140mm}
\vspace*{-80mm}
\caption{Behavior of the Energy density and entropy density as a function of temperature in Lattice QCD.}
\end{figure}

Recall that the  corresponding ideal value  $\epsilon^I/T^4$ reads:
\begin{equation}
\frac{\epsilon^I}{T^4}=(\nu_g +\frac{7}{8}\nu_q)\frac{\pi^2}{30}
\end{equation}

The behavior of the energy density ($\epsilon/T^4$) as a function of temperature ($T/T_c$) is shown in Fig.1 for pure gauge theory and for full QCD with $N_F=2,3$ in Fig.2. From these plots it is clear that $\epsilon/T^4$ approaches the ideal value only asymptotically. 

\subsection{The entropy-density}
We compute the entropy density for the interacting pure QCD as well the interacting quark-gloun plasma. This is again a straight forward exercise since we have the equilibrium distribution function for the quasi-partons already in hand. The entropy density in terms of  the grand canonical partition function reads:
\begin{eqnarray}
\label{eq6}
{\mathcal S}&=&\frac{1}{V}\partial_{T}[ T \ln({\bf{Z_g}})]+\frac{1}{V}\partial_{T} [T\ln({\bf{Z_M}})]\nonumber\\
\ln({\bf{Z_g}})&=&-V\nu_g \int \frac{d^3p}{2\pi^3}\ln(1-z_g\exp(-\beta p))\nonumber\\
\ln({\bf {Z_M}})&=& \ln({\bf {Z_q}})\nonumber\\
&&=V\nu_q\int \frac{d^3p}{2\pi^3}\ln(1+z_q\exp(-\beta p)).\nonumber\\
\end{eqnarray}

\begin{figure}[htb]
\label{fig7}
\vspace*{-70mm}
\hspace*{-40mm}
\psfig{figure=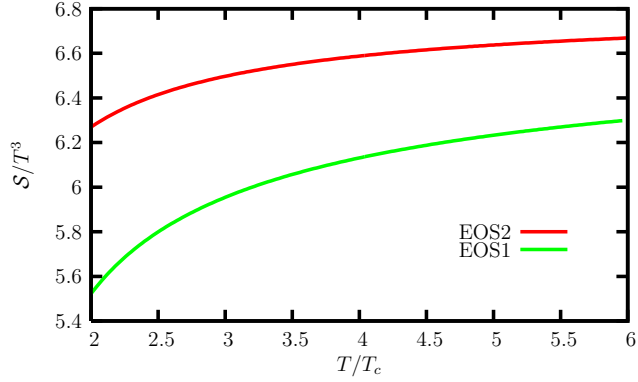,width=140mm}
\vspace*{-80mm}
\caption{Behavior of the entropy density  for pure gauge theory as a function of temperature}
\end{figure}

\begin{figure}[htb]
\label{fig8}
\vspace*{-70mm}
\hspace*{-40mm}
\psfig{figure=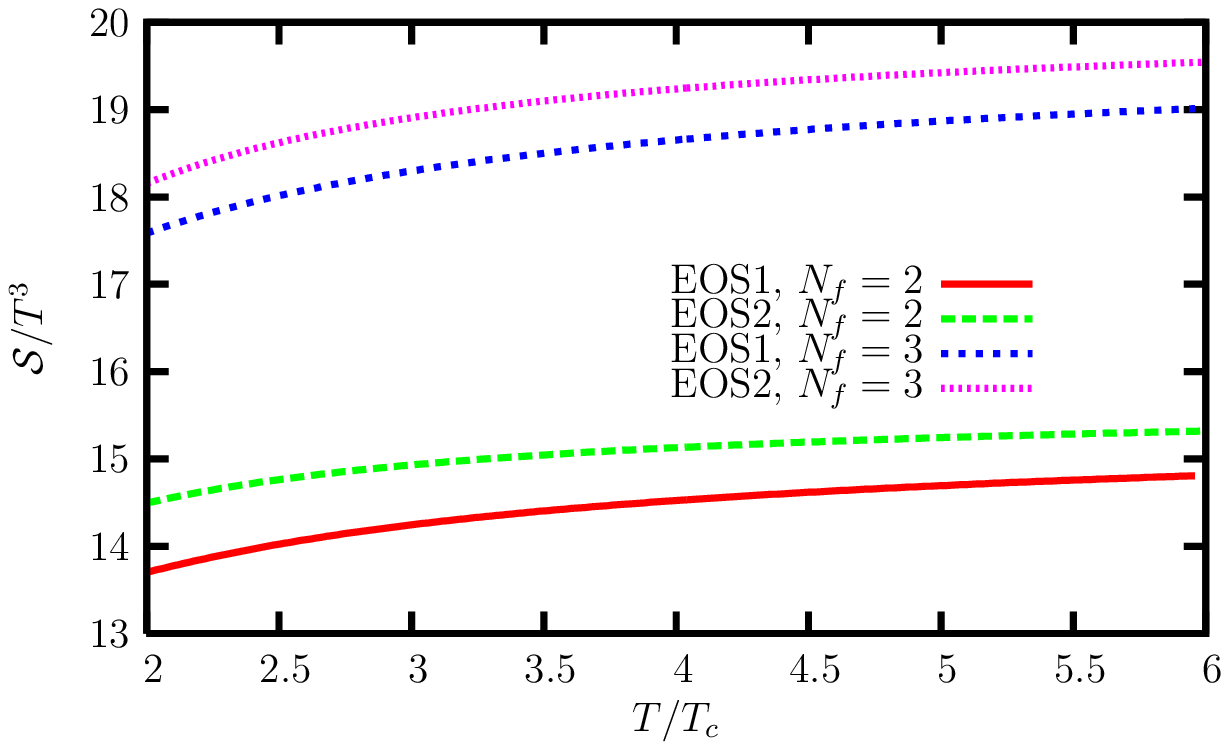,width=140mm}
\vspace*{-80mm}
\caption{Behavior of the entropy density in 2- and 3-flavor QCD as a function of temperature}
\end{figure}
The expression for the gluonic and quark contributions to the entropy density are obtained as,
\begin{eqnarray}
\label{eq7}
{\mathcal S_g}&=&\frac{\nu_g}{2\pi^2\beta^3}\bigg(8\ {\mathcal PolyLog}[4,z_g]\nonumber\\
 &&-2\tilde\mu_g {\mathcal PolyLog}[3,z_g]\bigg)+\Delta_g/T\nonumber\\
{\mathcal S_q}&=&\frac{\nu_q}{2\pi^2\beta^3}\bigg(-8\ {\mathcal PolyLog}[4,-z_q)]\nonumber\\
&& +2\tilde \mu_q {\mathcal PolyLog}[3,-z_q)]\bigg)+\Delta_q/T\nonumber\\
\end{eqnarray}

The total entropy can be obtained by adding the gluon and quark contributions,
${\mathcal S}={\mathcal S_g} + {\mathcal S_q}$. We have plotted the dimensionless quantity 
${\mathcal S}/T^3$ for pure gauge theory and full QCD $(N_f=2, 3)$, as a function of $T/T_c$ for 
EOS1 and EOS2. These are shown in  Figs.(4) and (5) respectively. We shall utilize the expression for entropy density  displayed
in Eq.\ref{eq7} for determining the viscosity to entropy density ratio.
 The corresponding ideal value for ${\mathcal S}/T^3$ is given by,
\begin{equation}
\frac{{\mathcal S}^I}{T^3}=(\nu_g +\frac{7}{8}\nu_q)\frac{2\pi^2}{45}.
\end{equation}

\subsection{Comparison with lattice results}
 We compare the thermodynamic observables determined by employing this model with  the lattice results obtained by  Karsch\cite{fkarsch}. The lattice results\footnote{We thank Frithjof Karsch for providing us the pure lattice gauge theory data. 
} are shown in Figs 3, and our results are displayed in Figs 1,2 and 4-5. For 2- and 3-flavor lattice results, we consider Fig.14 of Ref.\cite{fkarsch}. The agreement is overall 
good for EOS2,  particularly beyond $2.5 T_c$; the agreement is merely qualitative for EOS1. These findings are consistent with our earlier result  \cite{chandra2} that EOS2 predictions for
the temperature dependence of the Debye mass and the dissociation temperatures of heavy quark ground states  are in broad agreement with the lattice values.

\section{Viscosity of interacting quark-gluon plasma}
\subsection{A brief review}
The determination of viscosity is not as  straight forward an exercise as the determination of the  thermodynamic observables.
For, it requires modeling beyond the equilibrium properties, in terms
of the collision terms and other transport parameters, and also the nature of the perturbation to the equilibrium distribution. We note
here that for sQGP under consideration, the collision term is by no means easily determined since
perturbative results involving lowest order contributions are by no means guaranteed to be
reliable. A reliable non-perturbative collision term is even harder to obtain. We need to adopt methods that go beyond the determination of properties such as the energy density or the specific heat.

There are two ways to compute the transport parameters for QGP:(i) from quantum field theory by using the Kubo formula\cite{arnold,dtson} or (ii) from the semi-classical transport theory\cite{landau,asakawa,arnold,greiner}.
To model the QGP produced in heavy ion collisions employing semi-classical transport theory, one needs to employ the
Vlasov term which incorporates the dynamics of non-abelian color charges. We also need a
reliable collision term, which is, as we have pointed out,  difficult to determine. A collision term which has recently been computed
by Arnold, Moore and Yaffe\cite{arnold} by considering binary collisions at the tree level in the lowest order.The  collision term so obtained is then utilized to estimate the shear viscosity for QGP\cite{arnold,asakawa}. Further, Asakawa {\it et. al} \cite{asakawa} have  included the Vlasov term for an 
ensemble of turbulent color fields, and determined the anomalous shear viscosity;this determination 
does not require a collision term.This work generalizes the  earlier work of Dupree \cite{dupree} by including the nonabelian dynamics. Both these works are based on methods describe in \cite{landau}.   As mentioned earlier,  studies based on AdS/CFT 
correspondence\cite{dtson,buchel} employ the Kubo formula and predict a lower bound 
of $(1/{4\pi})$ for $\eta/\mathcal{S}$
for plasmas whose dynamics are governed by
class of strongly coupled gauge theories. This speculation is supported in some studies based on
transport theory \cite{greiner,abhijit}. 

In a recent work\cite{asprl,abhijit}, it has been  argued that one does not need to treat 
QGP as a strongly coupled plasma to understand  low viscosity. The authors  argue that the 
anomalous transport processes in the rapidly expanding QGP are actually responsible for the very low value of the 
shear viscosity, and  not the binary collisions. On the other hand in a very recent work, Xu and Greiner\cite{greiner} argue that
the the reason for a small value of $\eta/\mathcal{S}$ is mainly   due to  gluon bremsstrahlung
contributions to the collision term.

In the present paper, we adopt the approach of Asakawa {\it et al}\cite{asakawa} and determine the
contribution of both the collisional and the anomalous parts to the ratio $\eta/\mathcal{S}$.  
As in the earlier study \cite{asprl}, we find that the anomalous part dominates over the collisional contribution. 
 
\subsection{ Determination of $\eta$} 
In this Section, we determine the viscosity of a rapidly expanding interacting QGP, as described
by EOS1 and EOS2, and as represented by equivalent equilibrium distribution functions (section 2).
Our procedure involves replacing the ideal gas distributions used in \cite{asakawa}, by the
ones obtained by us for EOS1 and EOS2. We find  that all the assumptions made in Refs.\cite{asakawa,asprl}
will be applicable in the present case.

Let us first briefly outline the standard procedure of determining viscosity in transport theory\cite{landau,asakawa}.
The shear  viscosity of QGP in terms of parton occupation numbers can be obtained by comparing the 
microscopic definition of the stress tensor with the macroscopic definition of the viscous stress tensor. 
The microscopic definition of the stress tensor in terms of the distribution function is as follows:
\begin{equation}
\label{eq8}
 T_{i k}=\int \frac{d^3p}{(2\pi)^3 E_p} p_i p_k f(\vec{p},\vec{r}).
\end{equation}
On the other hand the macroscopic expression for the viscous stress tensor reads:
\begin{equation}
\label{eq9}
T_{i k}=P\delta_{i k}+\epsilon u_i u_k-2\eta(\nabla u)_{i k} -\zeta \delta_{i k} \nabla\cdot\vec{u},
\end{equation}
where $\eta$ is the shear viscosity, $\zeta$ is the bulk viscosity and $(\nabla u)_{i k}$ is the traceless, symmetrized
velocity gradient, 

\begin{equation}
\label{eq9b}
(\nabla u)_{i j}=\frac{1}{2} (\nabla_i u_j+\nabla_j u_i)-\frac{1}{3}\delta_{ij}\nabla\cdot \vec{u}.
\end{equation}

To obtain the shear viscosity, we write the distribution function as

\begin{equation}
\label{eqn9c}
f(\vec{p},\vec{r})=\frac{1}{z_{g/f}\exp(-\beta E_p+f_1(\vec{p},\vec{r}))\mp 1}.
\end{equation}
Assuming that  $f_1(\vec{p},\vec{r})$ is a small perturbation to the equilibrium distribution, we expand $f(\vec{p},\vec{r})$ and keep the linear order term in $f_1$; the following form of the  distribution function is thus obtained:

\begin{eqnarray}
 f(\vec{p},\vec{r})&=&f_0({\bf p})+\delta f(\vec{p},\vec{r})\nonumber\\
&=&f_0({\bf p})\bigg[1+f_1(\vec {p},\vec{r})(1\pm f_0({\bf p})\bigg],
\end{eqnarray}
 where $f_0\equiv f^{g/f}_{eq}$( see Eq.\ref{eq3a}).

The important point to be noted is that while deriving the transport  coefficients, one  assumes a slow
variation of the particle  distribution  so that the deviation from the equilibrium distribution is homogeneous in space  and proportional  to the gradients of the equilibrium parameters. Employing
the standard approach \cite{landau,asakawa}, we write 
\begin{equation}
\label{eq11}
 f_1(\vec{p},\vec{r})=-\frac{\bar{ \triangle}(p)}{E_p T^2} p_i p_j (\nabla u)_{i j},
\end{equation}
 where  the dimensionless function $\bar{ \triangle}(p)$  measures the deviation from the 
equilibrium configuration. 
Since $\eta$ is a Lorentz scalar, it may be
evaluated conveniently in the local rest frame. 
For  a boost invariant longitudinal flow, 
$(\nabla u)_{i j} = \frac{1}{3\tau} diag (-1, -1,2)$ in the local rest frame, and 
and  $f_1(p)$ takes the  form
\begin{equation}
\label{eq9a}
f_1(\vec{p})=-\frac{\bar{\triangle}(p)}{E_p T^2\tau}\bigg(p_z^2-\frac{p^2}{3}\bigg),
\end{equation}
where $\tau$ is the proper time($\tau=\sqrt{t^2-z^2}$).

The expression for $\eta$ is then obtained as 
\begin{equation}
\label{eq10}
\eta=\frac{-\beta}{15}\int \frac{d^3 p}{8\pi^3} \frac{p^4}{E_p^2} \bar{ \triangle}(p) \frac{\partial f_{eq}}{\partial E_p},
\end{equation}
entirely in terms of the unknown function $\bar{ \triangle}(p)$.
$E_p$ is the particle energy. 

We adopt  the ansatz in \cite{asakawa} and take the form of $ \bar{\triangle}(p)$ to be 
\begin{equation}
\label{eq13a}
 \bar{\triangle}(p)=A \vert p \vert/T; A\equiv{A_g,A_q}.  
\end{equation}

 To appreciate the ansatz better, we note that  the dispersion relation in Eq.{\ref{des}) gets modified, in the
presence of the perturbation  $f_1$ as given by
\begin{equation}
E_{eff}(p)=E_p-\frac{pA}{E_p T^3\tau}\bigg(p_z^2-\frac{p^2}{3}\bigg).
\end{equation}
 The velocity is given by 
 ($\vec{ v}_p=\partial_{\vec {p}} E_{eff}(p)$) 
\begin{eqnarray}
\vec{v}_p= \hat{p}-\frac{A}{T^3\tau}(p_z \hat{ k}-\frac{p}{3}\hat{p})
\end{eqnarray}

From this expression for $\vec{v}_p$, it is clear that the perturbation leads to different velocities in transverse and longitudinal directions. This introduces manifest anisotropy in the system.

  We determine $\bar{\triangle}(p)$ by the variational method by minimizing the linearized transport equation\cite{landau,asakawa} with a Vlasov term and a collision term computed by Arnold {\it et. al}\cite{arnold}.
 
 The factor $A$ is yet undetermined. To fix its value, we minimize the quadratic functional \cite{asprl,asakawa},
\begin{eqnarray}
\label{eq13}
W[\bar{f_1}]&=&\int \frac{d^3p}{8\pi^3} \bar{f_1(\vec{ p})}\bigg[v^{\mu} \partial_{x_\mu} f(\vec{p}) 
\nonumber\\ &&+\frac{1}{2}(-\nabla_p \cdot
D\cdot\nabla_p \delta \bar{f}(\vec{p}) + I[\bar{f_1}(\vec{p})\bigg],\nonumber\\
\end{eqnarray}
where the first term gives the drift, the second represents the diffusive Vlasov dynamics, and the last
term is the collision integral.
The expression in parenthesis is just the transport equation satisfied by $f_1(p)$ after averaging over
the color fields \cite{asakawa}. The  equilibrium distribution functions modify the results
obtained in \cite{asakawa} by rendering the coefficients $A_q,~A_g$ dependent on temperature and the coupling
constant. After performing the momentum integrals, the drift, diffusive and the collisional terms in the quadratic functional Eq.\ref{eq13} acquire
finally the form
\begin{eqnarray}
\label{eq23}
\tilde{W}_{D}[\bar{f_1}] &=& \frac{-32 \vert \nabla u\vert^2 }{3\pi^2} T^2\bigg[ (N_c^2-1)I^g_5\ A_g +N_c N_f \ I^q_5 A_q\bigg],\nonumber\\
\tilde{W}_{V}[\bar{f_1}] &=& \frac{16\vert \nabla u\vert^2}{5\pi^2 T} g^2 \langle B^2 \rangle \tau_m \bigg[ N_c\ I^g_4 \ A_g^2 + N_f I^q_4 \ A_q^2 \bigg],\nonumber\\
\tilde{W}_{C}[\bar{f_1}] &=& \frac{\vert \nabla u\vert^2 T^2}{2} (N_c^2-1)g^4\log(g^{-1})\nonumber\\
&&\times\bigg[\frac{7}{24\pi^2}(2N_c + N_f) \bigg(N_c \ \frac{I^g_2}{z_g} A^2_g \nonumber\\
&&+ N_f \ \frac{I^q_2}{z_q} A^2_q \bigg)\nonumber\\
&&+ \frac{ N_f N_c}{2\pi^3}(N_c^2-1)\bigg(z_g\frac{I^q_4 + I^g_4}{z^g+z^q}\bigg) (A_q -A_g)^2\bigg],
\end{eqnarray}
where
\begin{eqnarray}
\label{eq21}
I^g_n=PolyLog[n,z_g^{-1}]\nonumber\\
I^q_n=-PolyLog[n,-z_q^{-1}].\nonumber\\
\end{eqnarray}
The function $PolyLog[n,a]$ has the series representation
\begin{equation}
\label{eq22}
PolyLog[n,a]=\sum_{k=0}^{\infty} \frac{a^k}{k^n}.
\end{equation}
These expressions reduce to the ones obtained in \cite{asakawa}, if we put $z_q,~z_g=1$, corresponding
to ideal quark and gluon distributions.

\subsection{The anomalous and collisional viscosities}
Let us now turn our attention to determine the analytic expressions for
anomalous and collisional contributions to the shear viscosity, which are determined respectively by
the diffusive Vlasov and the collision terms in Eq.\ref{eq23} . To determine either of them, 
one utilizes  Eq.\ref{eq10} along with Eq.\ref{eq23} by following exactly the path taken in Ref\cite{asakawa}.

By inserting Eq.\ref{eq10} in Eq.\ref{eq13a} and performing the momentum integration,  we  obtain the following expression for viscosity $\eta$:
\begin{equation}
\label{eq24}
 \eta =\frac{8}{\pi^2} \beta^{-3}\bigg[(N_c^2-1)I^g_5 \ A_g + N_c N_f I^q_5 \ A_q\bigg].
\end{equation}

The minimization of the  functional $\tilde{W}[\bar{f_1}]$(Eq.\ref{eq23}) leads to the following matrix equation for the column vector $A = (A_g,~A_q)$:
\begin{equation}
\label{eq25}
\bigg(\tilde{a}_A + \tilde{a}_C\bigg)A= \tilde{r},
\end{equation}
where, the  column vector $\tilde{r}$ and the matrices $\tilde{a}_A$ and $\tilde{a}_C$ are  given by
\begin{equation}
\label{eq26}
\tilde{r} = \frac{32}{3\pi^2}\left(\begin{array}{c}
(N_c^2-1)I^g_5 \\
N_cN_f I^q_5
\end{array}
\right)
\end{equation}

\begin{equation}
\label{eq26a}
\tilde {a}_{A} = \frac{32}{5\pi^2}
\frac{g^2 \langle{\cal B}^2 \rangle\, \tau_{ m}}{T^3}
\left(
\begin{array}{cc}
N_cI^g_4 & 0 \\
0 & N_f I^q_4
\end{array}
\right)
\end{equation}

\begin{eqnarray}
\label{eq26b}
\tilde{a}_{C} &=&
\frac{7}{24\pi^2}(2N_c+N_f){\bf C_g}\left(\begin{array}{cc}
N_c \frac{I^g_2}{z_g}& 0 \\
0 & \frac{I^q_2}{z_q}N_f 
\end{array}\right)\nonumber\\
&&+ \frac{N_fN_c(N_c^2-1)}{2\pi^3}{\bf C_g}\frac{z_g}{z_q+z_g}\nonumber\\
&&\times(I^q_4 + I^g_4) \left(
\begin{array}{cc}
1 & -1 \\
-1 & 1 
\end{array}\right) \,
\end{eqnarray}
where ${\bf C_g}=(N_c^2-1)g^4\log( g^{-1})$.
This leads to the following expressions for the anomalous and collisional contributions to the
shear viscosity,
\begin{eqnarray}
\label{eq27}
\tilde{\eta}_{A} =\frac{3}{4} \beta^{-3} \tilde{r} \cdot (\tilde a_{A}^{-1})\cdot \tilde{r}\nonumber\\
\tilde{\eta}_{C} =\frac{3}{4} \beta^{-3} \tilde{r}\cdot (\tilde a_{C}^{-1})\cdot \tilde{ r}
\end{eqnarray}

By employing  the additivity of rates,  the expression for the total viscosity is obtained as \cite{asprl}
\begin{equation}
\label{eq27a}
\frac{1}{\eta}=\frac{1}{\eta_c}+\frac{1}{\eta_A}.
\end{equation}

It is clear from Eqs.\ref{eq26},\ref{eq26a}, \ref{eq26b} and \ref{eq27} that $\eta_c\sim\frac{1}{ g^4\ln(1/g)}$. 
In the weak coupling limit $(g<<1)$, at which the hot EOS are also valid,  the total viscosity in Eq.\ref{eq27a} will be 
dominated by the anomalous component. Therefore for a weakly coupled QGP $\eta \approx \eta_A$. Hence we
confine our attention to 
anomalous shear viscosity and  study it's behavior with temperature. The individual expressions for the gluon and quark contribution to the 
anomalous viscosity from Eq.\ref{eq27} are obtained to be

\begin{eqnarray}
\label{eq28}
\eta^g_A &=&\frac{40\beta^{-6}}{3\pi^2 g^2\langle B^2\rangle \tau_m}
\frac{(N_c^2-1)^2 (I^g_5)^{2}}{N_c I^g_4}\nonumber\\
\eta^q_A &=&\frac{40\beta^{-6}}{3\pi^2 g^2\langle B^2\rangle \tau_m}
\frac{N_c^2(I^q_5)^{2}}{N_f I^q_4}.
\end{eqnarray}
 The total anomalous shear viscosity is obtained by summing up these two contributions,
({\it i.e}, $\eta_A=\eta^g_A+\eta^q_A$). We note that the above expressions are valid for a  purely magnetic plasma. 
For the case when both chromo-electric and chromo -magnetic fields are  present in the turbulent phase,  and all their components 
are of equal size, the expressions for viscosity can be obtained simply by the replacement $<B^2>\tau_m\rightarrow \frac{4}{3}(<{\mathcal E}^2>+<B^2>)\tau_m$\cite{asakawa}. 

Accordingly, we  rewrite Eq.\ref{eq28} as
\begin{eqnarray}
\label{eq29}
\eta^g_A(z_g) &=&\frac{10\beta^{-6}}{\pi^2 g^2\langle {\mathcal E}^2+ B^2\rangle \tau_m}
\frac{(N_c^2-1)^2 (I^g_5)^{2}}{N_c I^g_4}\nonumber\\
\eta^q_A(z_q) &=&\frac{10\beta^{-6}}{\pi^2 g^2\langle{\mathcal E}^2+ B^2\rangle \tau_m}
\frac{N_c^2(I^q_5)^{2}}{N_f I^q_4}.
\end{eqnarray}

We pause to compare the viscosities with their ideal values. Recall that the contribution from ideal
distribution functions are obtained by setting
 $z_g=1$ and $z_q=1$ in Eqs.\ref{eq28} and \ref{eq29};  
as expected they match with the expressions of Asakawa {\it et. al}\cite{asakawa}. Thus, the expressions for the relative viscosities (anomalous) are read off as

\begin{eqnarray}
\label{eq32}
{\mathcal R}_g\equiv\frac{\eta^g_A}{\eta^{I,g}_A}&=&\frac{\zeta(4)}{\zeta(5)^2} \frac{(I^g_{5})^2}{I^g_4}\nonumber\\
{\mathcal R}_q\equiv\frac{\eta^q_A(z_q)}{\eta^{I,q}_A}&=&\frac{56\zeta(4)}{225 \zeta(5)^2} \frac{(I^q_{5})^2}{I^q_4}.
\end{eqnarray}
\begin{figure}[htb]
\label{fig9}
\vspace*{-70mm}
\hspace*{-40mm}
\psfig{figure=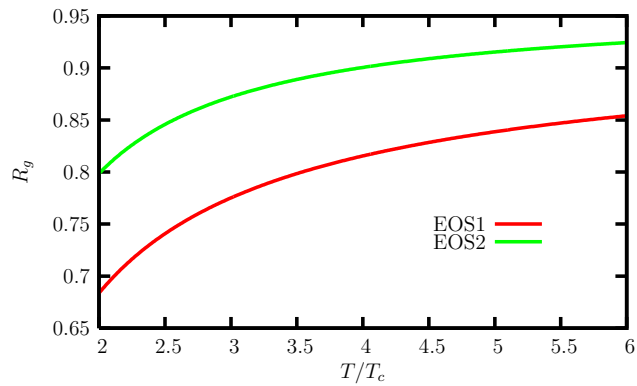,width=140mm}
\vspace*{-80mm}
\caption{Behavior of the relative viscosity for pure gauge theory as a function of temperature}
\end{figure}
\begin{figure}[htb]
\label{fig10}
\vspace*{-70mm}
\hspace*{-40mm}
\psfig{figure=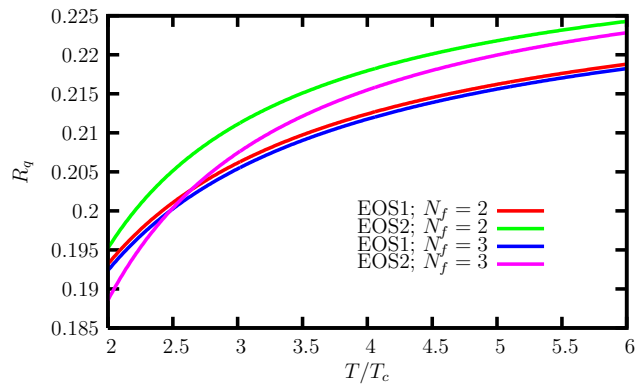,width=140mm}
\vspace*{-80mm}
\caption{Behavior of the quark contribution to the viscosity in 2- and 3-flavor QCD as a function of temperature}
\end{figure}
 The behavior of ${\mathcal R}_g$ and ${\mathcal R}_q$ as a function of temperature for EOS1 and EOS2 are shown in Figs.(7) and (8). Clearly incorporation of interaction effects in the EOS further reduces the
viscosities. While the gluon viscosity can reduce upto $\sim 30 \%$, the fall in the quark viscosity can be as steep as 80\%, indicating a more ideal fluid like behaviour.

\subsection{Viscosity to entropy density ratio}
A determination of the absolute values of viscosity requires further a knowledge of the quenching
parameter $\hat{q}_R$\cite{abhijit}, which is defined as the rate of growth of the transverse momentum fluctuation of a fast parton in an ensemble of turbulent color fields \cite{abhijit}. In turn, $\hat{q}_R$ is given by
\begin{equation}
\label{eq30}
 \hat{ q}_R = \frac{8\pi\alpha_s N_c}{3(N_c^2-1)}\langle E^2 + B^2\rangle \tau_m,
\end{equation}
in terms of the total energy density and  an appropriate relaxation time $\tau$ \cite{asakawa}.

 One can combine this with the expression for the anomalous viscosity of gluons and obtain the relation as follows,
\begin{equation}
\label{eq31}
 \eta^g_A(z_g)=\frac{20 T^6_c}{3\pi^2 \hat q_R}(N_c^2-1) (\frac{T}{T_c})^6 \frac{(I^g_5)^{2}}{ I^g_4}.
\end{equation}
Estimates for $\hat{q}_R$ are available for the gluonic case only. In this case, 
 $\hat{q}_R$ is estimated  from the data by various approaches\cite{maju}. Studies within the framework of the twist expansion by fitting the experimental data on hadron suppression in the most central Au-Au collisions \cite{zhang,maju1} yield   values in the range $1-2 GeV^2/fm$ for the gluon quenching parameter. On the other hand, an  eikonal approach\cite{schi,acsw}
estimates it to be roughly ten times larger than the twist estimates, in the range $10-30 GeV^2/fm$. 
The expression for the gluonic $\eta/{\mathcal S}$ is obtained as
\begin{equation}
\label{eqs}
\frac{\eta}{{\mathcal S}}=\frac{20 T_c^3(N_c^2-1)}{3\hat{q}_R}(\frac{T}{T_c})^3 \frac{(I^g_5)^2}{I^4_g[\nu_g(4I^g_4-\ln(z_g) I^g_3)+\frac{\pi^2\Delta_g}{T^4}]}.
\end{equation}
It is clear that the estimates for $\eta/{\mathcal S}$  inherit the uncertainty in $\hat{q}_R$,  upto an order of
magnitude.  For purposes of concreteness, we choose the QCD transition temperature($T_c$) to be $270 Mev$\cite{zantow}.  
We have plotted Eq.\ref{eqs} as function of temperature for EOS1 and EOS2 in Fig.8
 and 9, with respective values $\hat{q}_R = 1,~10 GeV^2/fm$. 
 We see that in the latter case, the ratio can fall
significantly below the AdS/CFT bound  $\frac{1}{4\pi} \sim 0.08$ even at $3T_c$, but it may not be reliable since the large  value of (${\hat q}_R=10 GeV^2/fm$) which we have employed  may not be accomodated within 
weak coupling framework\cite{abhijit} which we consider in the present paper.  On the other hand, in the former case ${\hat q}_R=1 GeV^2/fm$ the value of $\eta/{\mathcal S}$ does not violate   the AdS/CFT bound, although it is quite close to it near $2T_c$. It appears that the violation of bound, which can occur at ${\hat q}_R >| 1 GeV^2/fm$
will be marginal near $2T_c$. 

 As expected the ratio increases with increasing temperature. Interestingly, unlike other thermodynamic variables, the ratio
is not sensitive to the EOS employed.
\begin{figure}[htb]
\label{fig11}
\vspace*{-70mm}
\hspace*{-40mm}
\psfig{figure=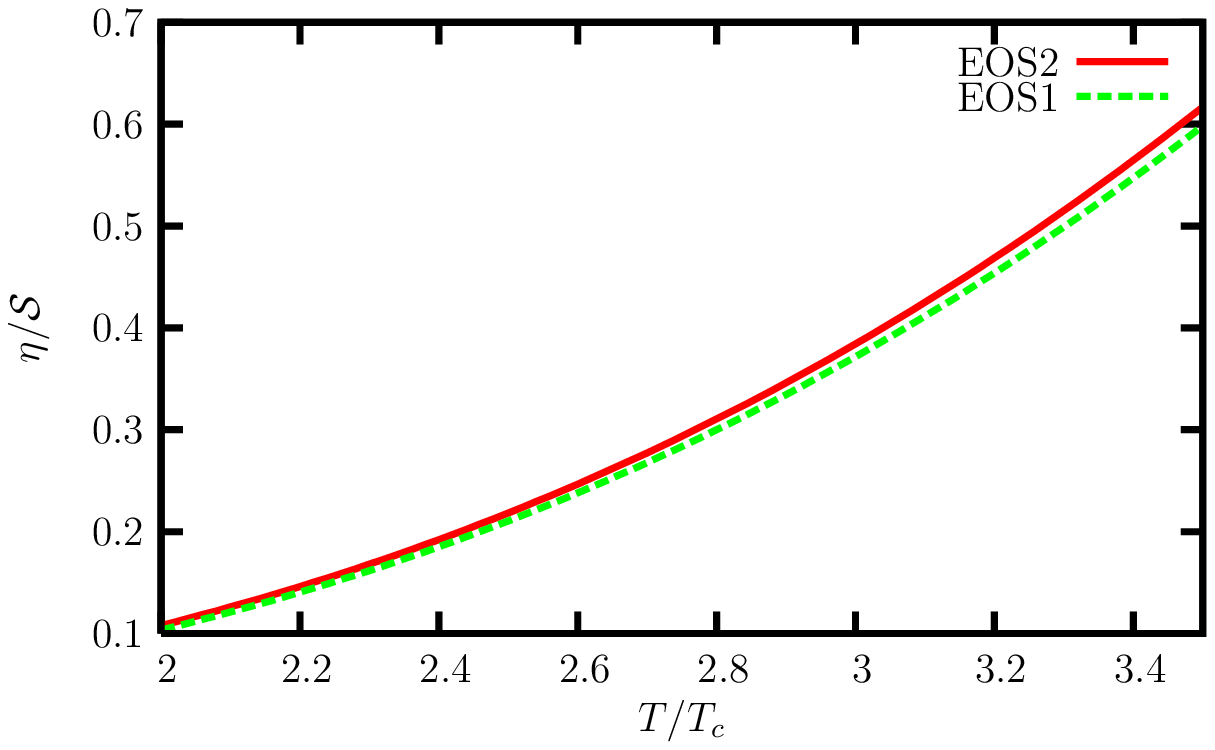,width=140mm}
\vspace*{-80mm}
\caption{Viscosity to entropy density ratio for pure gauge theory for $\hat{q_R}=1 GeV^2/{fm}$}
\end{figure}

We now establish the connection between our results and that of Asakawa {\it et al} \cite{asakawa}.
The expression for the ratio $\eta/{\mathcal S}$ in Eq.\ref{eqs}  reduces to that of Asakawa {\it et al} if we set $z_g =1 $ and  employ the ansatz, $\bar \Delta(p)=A_{g/q} p/T$ for the anisotropy parameter \footnote{ In writing Eq.\ref{eqs1}, 
we  have employed the following results: ${\mathcal PolyLog}[4,z_g=1]=\zeta(4)=\pi^4/90$ and ${\mathcal PolyLog}[5,z_g=1]=\zeta(5)$ and that $S^I_g=4\nu_g\zeta(4)T^3/\pi^2$. Eq.\ref{eqs1} follows from from Eqs.(6.32 and 6.33) in Ref.\cite{asakawa} by employig the  form of  ${\mathcal S}^I_g$ given above.}. The corresponding expression in this limit reads,
\begin{equation}
\label{eqs1}
\frac{\eta}{{\mathcal S}}=\frac{20 T_c^3(N_c^2-1)}{\nu_g \hat{q}_R}(\frac{T}{T_c})^3 \frac{\zeta(5)^2}{4\zeta(4)^2},
\end{equation}

 Let us normalize the viscosity to entropy ratios for EOS1 and EOS2 {\it wrt} the ideal values. We have plotted the relative ratios.   which we denote by ${\mathcal R}_\eta$,   as a function of temperature,  in Fig.10, which shows the effects of interactions in $\eta/{\mathcal S}$. From Fig.10, we see that (${\mathcal R}_\eta$) is
less than unity, approaching the ideal value asymptotically. Interestingly, the EOS2 values are closer to the ideal case, differing by about $3\%$ 
near $2T_c$.

 Finally, note that the expression for $\eta/{\mathcal S}$ in Eq.\ref{eqs1} is 
identical  to  the expression used in \cite{abhijit} except for   a numerical factor of $O(1)$. 
This discrepancy arises because we consider both the diffusive Vlasov and the collision terms in the transport 
equation, while the analysis of \cite{abhijit} neglects the collision term.

\begin{figure}[htb]
\label{fig12}
\vspace*{-70mm}
\hspace*{-40mm}
\psfig{figure=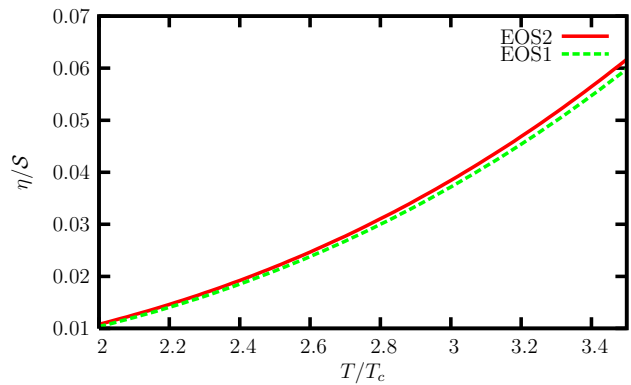,width=140mm}
\vspace*{-80mm}
\caption{Viscosity to entropy density ratio for pure gauge theory for $\hat{q_R}=10 GeV^2/{fm}$}
\end{figure}

\begin{figure}[htb]
\label{fig13}
\vspace*{-70mm}
\hspace*{-40mm}
\psfig{figure=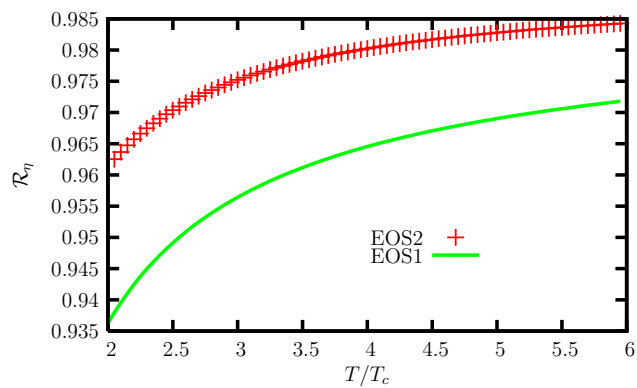,width=140mm}
\vspace*{-80mm}
\caption{Behavior of ${\mathcal R}_{\eta}$ as a function of temperature in the case of pure 
gauge theory for EOS1 and EOS2. Note that ${\mathcal R}_{\eta}$ scales with $T/T_c$.}
\end{figure}

\section{Conclusions and outlook}
In conclusion, we find that hot QCD EOS corresponding to interactions of $O(g^5)$ and $O(g^6\ln(1/g)$
can significantly impact the values of the thermodynamic observables such as the energy density.
The viscosity and the ratio $\eta/{\mathcal S}$, which we have studied as functions of
temperature, get reduced by approximately $7\%$ for EOS1 and $4\%$ for EOS2 near $2T_c$ in contrast to their ideal 
counterparts.
We found that the value of $\eta/{\mathcal S}$  for ${\hat q}_R=1 GeV^2/fm$ near $2T_c$ is closer to the lower bound $1/{4\pi}$ placed on $\eta/{\mathcal S}$ by AdS/CFT studies.  Further, the choice  ${\hat q}_R\sim 10 GeV^2/fm$  is difficult to accommodate within the weak perturbative framework and hence the violation of the AdS/CFT bound may not represent the factual situation. The choice ${\hat q}_R\sim 2 GeV^2/fm$ does lead to a violation near $2T_c$, but only marginally so. In short,   the findings in the present work strengthen
the near perfect fluid picture of the hot and dense matter created in relativistic heavy ion collisions.
This analysis has
been rendered possible because of the mapping of interacting partons to non-interacting
quasi partons with effective fugacities \cite{chandra1,chandra2}. While the present study points definitively
to the importance of interaction effects, it is by no means complete, because of inherent uncertainties in the estimates of the gluonic quenching parameter, and an absence of the knowledge of the quenching parameter
for the quarks. The EOSs which we study are also perturbative. It should be of great interest to employ the lattice EOS\cite{fkarsch,lattice}. This will be taken up separately.

\vspace{3mm}
\noindent{\bf Acknowledgment:}\\
We are  thankful to Frithjof Karsch for providing us lattice data for pure gauge theory and Anton Rebhan for useful comments and suggestions on fixing $O(g^6)$ contributions at $O(g^6\ln(1/g))$.  VC thanks  Raman Research Institute for  hospitality where the manuscript was finalized and CSIR, New Delhi(India) for financial support.

\end{document}